\documentclass[aps,prl,reprint,superscriptaddress]{revtex4-2}
\usepackage{graphicx}
\usepackage{amsmath}
\usepackage{mathcomp}
\usepackage{textcomp}
\usepackage{siunitx}

\begin{document}

%\preprint{}

\title{An integrated magnetometry platform with stackable waveguide-assisted detection channels for sensing arrays}


\author{Michael Hoese}
\email[M.H. and M.K.K. contributed equally to this work.]{}
\affiliation{Institute for Quantum Optics, Ulm University, D-89081 Ulm, Germany}

\author{Michael K. Koch}
\email[M.H. and M.K.K. contributed equally to this work.]{}
\affiliation{Institute for Quantum Optics, Ulm University, D-89081 Ulm, Germany}
\affiliation{Center for Integrated Quantum Science and Technology (IQst), Ulm University, D-89081 Ulm, Germany}

\author{Vibhav Bharadwaj}
\affiliation{Institute for Photonics and Nanotechnologies (IFN) - CNR, Piazza
Leonardo da Vinci, 32, Milano, 20133, Italy}

\author{Johannes Lang}
\affiliation{Institute for Quantum Optics, Ulm University, D-89081 Ulm, Germany}

\author{John P. Hadden}
\affiliation{School of Physics and Astronomy, Cardiff University, Cardiff CF24 3AA, United Kingdom}

\author{Reina Yoshizaki}
\affiliation{Department of Mechanical Engineering, School of Engineering, The
University of Tokyo, Tokyo, 113-8656, Japan}

\author{Argyro N. Giakoumaki}
\affiliation{Institute for Photonics and Nanotechnologies (IFN) - CNR, Piazza
Leonardo da Vinci, 32, Milano, 20133, Italy}

\author{Roberta Ramponi}
\affiliation{Institute for Photonics and Nanotechnologies (IFN) - CNR, Piazza
Leonardo da Vinci, 32, Milano, 20133, Italy}

\author{Fedor Jelezko}
\affiliation{Institute for Quantum Optics, Ulm University, D-89081 Ulm, Germany}
\affiliation{Center for Integrated Quantum Science and Technology (IQst), Ulm University, D-89081 Ulm, Germany}

\author{Shane M. Eaton}
\affiliation{Institute for Photonics and Nanotechnologies (IFN) - CNR, Piazza
Leonardo da Vinci, 32, Milano, 20133, Italy}

\author{Alexander Kubanek}
\email[Corresponding author: ]{alexander.kubanek@uni-ulm.de}
\affiliation{Institute for Quantum Optics, Ulm University, D-89081 Ulm, Germany}
\affiliation{Center for Integrated Quantum Science and Technology (IQst), Ulm University, D-89081 Ulm, Germany}

\date{\today}

\begin{abstract}

The negatively-charged NV$^-$-center in diamond has shown great success in nanoscale, high-sensitivity magnetometry. Efficient fluorescence detection is crucial for improving the sensitivity. Furthermore, integrated devices enable practicable sensors. Here, we present a novel architecture which allows us to create NV$^-$-centers a few nanometers below the diamond surface, and at the same time in the mode field maximum of femtosecond-laser-written type-II waveguides. We experimentally verify the coupling efficiency, showcase the detection of magnetic resonance signals through the waveguides and perform first proof-of-principle experiments in magnetic field and temperature sensing. The sensing task can be operated via the waveguide without direct light illumination through the sample, which marks an important step for magnetometry in biological systems which are fragile to light. In the future, our approach will enable the development of two-dimensional sensing arrays facilitating spatially and temporally correlated magnetometry.

\end{abstract}

%\keywords{}

%\maketitle must follow title, authors, abstract, and keywordk
\maketitle


\section{Introduction}

Quantum sensing performed with negatively-charged nitrogen vacancy centers (NV$^-$-centers) in diamond has successfully measured strain \cite{Doherty2014}, temperature \cite{Plakhotnik2014} and magnetic fields \cite{Maze2008, Taylor2008, Balasubramanian2008} at the nanoscale and with high sensitivity. NV$^-$-based magnetometry is widely applied to characterize biological samples such as cells \cite{Kucsko2013} or novel materials \cite{Kolkowitz2015, Lovchinsky2017}. The combination with atomic-force microscopy has revolutionized scanning-probe magnetometry \cite{Degen2008, Maletinsky2012} with applications in material science and life sciences. Advanced photonics could further improve the performance in terms of losses, signal-to-noise and operation speed \cite{Atature2018, Awschalom2018} and includes integration into optical fibers \cite{Ruan2018, Bai2020}, diamond nanopillar arrays \cite{Hanlon2020, Babinec2010} and integrated photonics \cite{Englund2010, Fehler2019a, Olthaus2020}. The latter enables, in addition, the realization of compact devices. In this context, femtosecond-laser-writing is an outstanding fabrication method that does not rely on lithography steps and has, in particular, the capability to fabricate three-dimensional structures. Using diamond as photonics host offers the additional advantage that NV$^-$-centers can directly be integrated into the photonic device. Laser-written, type-II waveguides in diamond have recently been developed \cite{Sotillo2016, Eaton2019} and show great potential to efficiently interface NV$^-$-centers. Thereby, a focused laser beam creates two nearby lines of reduced refractive index, where the stressed region between the two lines serves as waveguide. The controlled creation of defect centers has been demonstrated with ion implantation \cite{Schroeder2017} and laser-writing \cite{Chen2017, Hadden2018, Chen2019}. However, until today there is no strategy that enables deterministic post-processing of each waveguide in a three-dimensional platform, a problem that becomes even more challenging when color centers are required close to the diamond surface as it is the case for NV$^-$-based magnetometry.

\begin{figure*}[]
\includegraphics[scale=1.0]{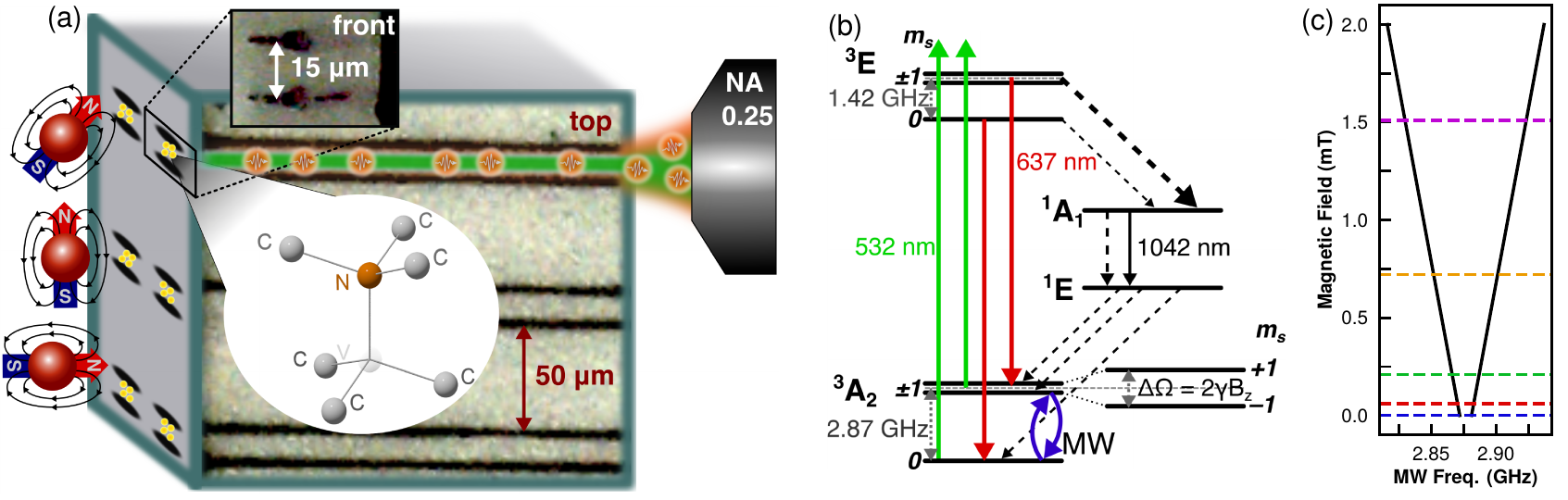}
\caption{\label{Fig:WaveguidePreparation}(a) Sketch of the waveguide-assisted sensor array. Shallow-implanted NV$^-$-centers act as local magnetometers on the diamond surface. The NV$^-$-centers are optically accessed by means of laser-written type-II waveguides and low-NA optics. The architecture can be extended to sensor arrays. The sketch incorporates microscope images of the waveguides in top and front view.
(b) Level scheme and dynamics of the NV$^-$-center as utilized to perform ODMR spectroscopy. Dashed arrows depict phonon transitions and bold arrows photon transitions, respectively. The arrow width reflects the interaction strength.
(c) Model for the splitting of the $m_s = \pm 1$ ground states with increasing magnetic field. The dashed lines indicate the magnetic field strengths that we apply in the experiments to perform ODMR scans.}
\end{figure*}

Here, we present a novel approach that is capable of functionalizing each waveguide in a three-dimensional architecture individually with NV$^-$-centers as depicted in figure \ref{Fig:WaveguidePreparation}(a). After laser-writing the waveguides, we create NV$^-$-centers through shallow-implantation of nitrogen ions on the front facet of the diamond photonics platform. The implantation depth of a few nanometers below the diamond surface enables the combination of efficient photon routing through diamond waveguides with sensing tasks on the diamond surface. Our architecture separates the optical access to the NV$^-$-centers from the object to be sensed by excitation and detection through the waveguides. Therefore, good sensing performance can be achieved without the need of light exposure through the sample. Ultimately, the architecture can be extended into a two-dimensional sensing array with potential for time-resolved and spatially-correlated magnetometry in material and life sciences. 

In the following, we discuss the working principle of our sensing device and characterize the waveguide-assisted optical access. Then, we perform first proof-of-principle magnetic field and temperature sensing with ensembles of NV$^-$-centers. We compare the sensor performance with respect to the state-of-the-art.
 
\section{Working principle}

Type-II waveguides are written in a CVD grown electronic grade diamond slab $\left(\SI{2}{\milli\metre}\times\SI{2}{\milli\metre}\times\SI{0.3}{\milli\metre}\right)$ with pulsed laser illumination at $\SI{515}{\nano\metre}$, with a repetition rate of $\SI{500}{\kilo\hertz}$, pulse width of $\SI{300}{\femto\second}$ and laser power of $\SI{100}{\milli\watt}$. The laser beam is focused into the sample through a high-NA objective (1.25NA, $100\times$) to create type-II waveguides of $\SI{2}{\milli\metre}$ length, according to the size of the diamond, in depths ranging from $\SI{5}{\micro\metre}$ to $\SI{25}{\micro\metre}$ below the top diamond surface. The waveguide depths are measured from the surface to the center of the modification. The type-II waveguide width of $\SI{15}{\micro\metre}$  (center to center transverse spacing between the two laser modification tracks as shown in fig. \ref{Fig:WaveguidePreparation}(a) front view) is optimized for single-mode, low-loss light transmission between $\SI{630}{\nano\metre}$ and $\SI{740}{\nano\metre}$ according to the NV$^-$-center sideband emission. In order to functionalize the waveguides with NV$^-$-centers, we shallow implant Nitrogen ions into the front facet of the waveguides followed by subsequent annealing at $\SI{1000}{\celsius}$. Further details on the fabrication can be found in the methods section.

The NV$^-$-centers spin-state, a spin-1 system in its ground and excited states, can be detected optically via magnetic resonance spectroscopy (ODMR) \cite{Jelezko2004}, as sketched in figure \ref{Fig:WaveguidePreparation}(b). Shelving from $m_s = \pm 1$ states induces optical pumping and allows to read out the state via measurement of the fluorescence intensity. The microwave (MW) resonance frequency decreases from $\SI{2.88}{\giga\hertz}$ at cryogenic temperatures to $\SI{2.87}{\giga\hertz}$ at room temperature thereby enabling, in principle, temperature sensing over a large temperature range although the sensitivity decreases at low temperatures. In a magnetic field the degeneracy of the $m_s = \pm 1$ states lifts due to a magnetic field dependent splitting of $\Delta\Omega = 2\gamma B_z$, with the gyromagnetic ratio $\gamma = \SI{28}{\giga\hertz\per\tesla}$, of the two states. As a result, the ODMR resonance dip splits into two, one for each spin state, thus enabling inference of the magnetic field strength. Figure \ref{Fig:WaveguidePreparation}(c) illustrates the magnetic field dependent splitting. Here, the degeneracy is already lifted at zero-field due to residual strain in the diamond lattice \cite{Kubo2010, Doherty2012}. For continuous ODMR (CW-ODMR) measurements, the overall sensitivity to DC magnetic fields \cite{Budker2007}, 
\begin{equation}
\eta_\text{DC} = \frac{\hbar}{g_\text{e} \mu_\text{B} \sqrt{N \tau}},
\end{equation}
depends on the number of NV$^-$-centers $N$ with coherence time $\tau$. Thereby, $g_\text{e} = 2.0$ denotes the electron $g$-factor, $\mu_\text{B}$ the Bohr magneton and $\hbar$ the Planck constant. One way to improve the overall sensitivity is to increase $N$. This can either be done by increasing the density of NV$^-$-centers or by enlarging the sensing volume. However, progressing to large sensing volumes remains challenging. The efficient extraction of photons out of the diamond host from a large ensemble of NV$^-$-centers distributed over a large sensing area is an outstanding problem. We therefore begin with the benchmarking of the device detection efficiency with respect to conventional, well-established confocal ODMR-measurements. We then extrapolate the overall sensitivity of the device by taking into account the large NV$^-$ ensemble that is addressable via the waveguide mode.  

\section{Waveguide-assisted optical access}

\begin{figure*}[]
\includegraphics[scale=1.0]{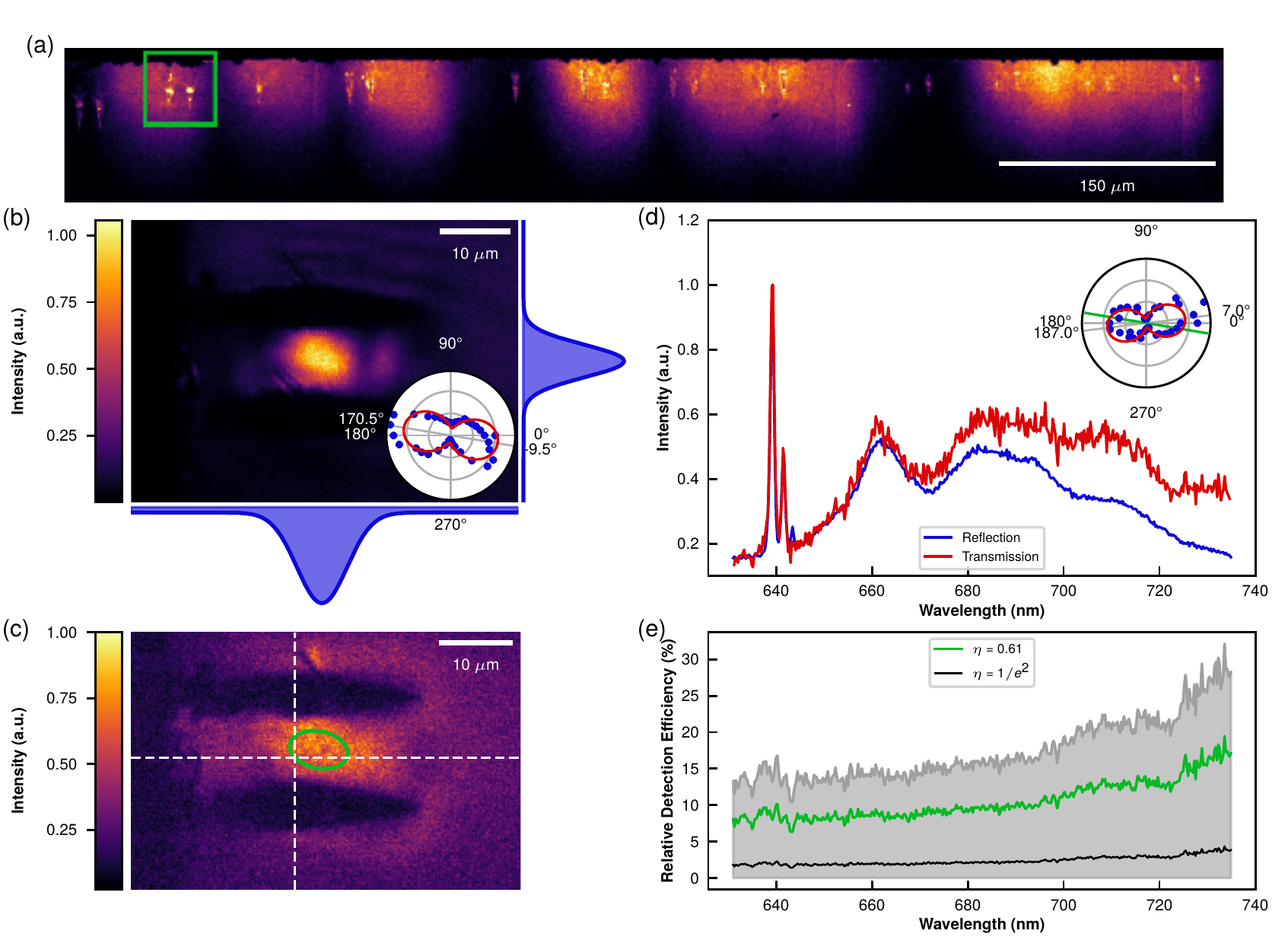}
\caption{\label{Fig:Fabrication}
(a) Confocal scan showing the fluorescence spots of the shallow-implanted NV$^-$-centers on the front facet of the sample. The leftmost waveguide (marked with a green box) is studied in the following.
(b) Mode shape of the waveguide (WG). The shape of the waveguide mode is revealed by laterally scanning a $\SI{738}{\nano\metre}$ laser while recording the collected counts in transmission. The two intersections at the side show the corresponding 2D-Gaussian fit to the WG mode. The corresponding mode-polarization contrast is shown in the lower right corner. Note that the image is rotated by \SI{90}{\degree} with respect to panel (a).
(c) NV$^-$ transmission through the waveguide. We excite the NV$^-$-centers off-resonantly with $\SI{532}{\nano\metre}$ laser light from the front and collect the transmitted signal. Laterally scanning the excitation laser resembles the waveguide mode in transmission. The intersection of the dashed lines denotes the position which is at $61 \%$ of the maximum of the WG mode (green ellipse). 
(d) The PL spectra in reflection and transmission are compared. The exposure time of the transmission measurement is ten times larger than for the measurement in reflection. Both spectra are normalized to the ZPL. The inset shows the ZPL polarization measured in transmission through the waveguide, which resembles the waveguide mode polarization (marked as green line). 
(e) We extract the wavelength specific optical detection efficiency of the NV$^-$ ensemble to the waveguide mode by comparing the transmitted PL spectrum with the fluorescence in reflection and by accounting for losses that occur until detection. The green curve is the inferred efficiency for a NV$^-$ ensemble located at $\SI{61}{\percent}$ of the optimal coupling corresponding to the green ellipse in (c). The black lines marks the $1/e^2$-profile of the Gaussian mode.
}
%\label{fig_WG_mode_polarization}
\end{figure*}

We first compare the detected signal through the waveguide with the confocal detection. Therefore, we excite a confocal spot of the NV$^-$ ensemble and simultaneously measure the transmission signal through the waveguide and the confocal signal in reflection. We examine the transmission properties for waveguides at a depth of $\SI{20}{\micro\metre}$ below the diamond surface with well-isolated waveguide modes. A confocal scan of the waveguide front facet, depicted in figure \ref{Fig:Fabrication}(a), shows the bright areas where NV$^-$-centers have been created. The areas around waveguide walls appear brighter indicating an increased number of NV$^-$-centers. In this study, we focus on the leftmost waveguide (green box). To characterize the transmission of the waveguide, we map the mode-excitation shown in figure \ref{Fig:Fabrication}(b) by laterally scanning a $\SI{738}{\nano\metre}$ laser while recording the maximum transmission signal. The 2D-Gaussian mode profile is projected onto the horizontal and vertical coordinate axes. We obtain a similar mode profile when exciting the NV$^-$ centers with $\SI{532}{\nano\metre}$ laser light and collecting the fluorescence in transmission through the waveguide while laterally scanning the excitation laser (see figure \ref{Fig:Fabrication}(c)). We note that the transmitted fluorescence resembles the mode profile, although the front facet is homogeneously covered with NV$^-$-centers (see confocal scan in fig. \ref{Fig:Fabrication}(a)). We conclude that NV$^-$ emission is guided through the waveguide.

In order to benchmark the detection efficiency of the waveguide-assisted optical access, we first compare the photoluminescence (PL) spectrum from the NV$^-$ ensemble in confocal configuration with the spectrum acquired in transmission through the waveguide as shown in figure \ref{Fig:Fabrication}(d). Note that in order to clearly determine the  NV$^-$ emission we cooled the sample to cryogenic temperatures. The resulting PL spectra, both in transmission and reflection, reveal the characteristic zero-phonon line (ZPL) and phonon-sideband (PSB) emission of NV$^-$-centers. Here, intensities are normalized with respect to the ZPL emission for sake of clarity. The first thing to notice is that the polarization pattern of the waveguide mode, shown in the inset of figure \ref{Fig:Fabrication}(b), yields a distinct polarization contrast which matches the polarization of the waveguided NV$^-$fluorescence within $\approx \SI{16.5}{\degree}$. From the measured transmission and reflection spectra we extract the ODMR-signal detection efficiency of our waveguide-assisted sensor with respect to conventional confocal measurements.

Figure \ref{Fig:Fabrication}(e) illustrates the ratio of transmitted signal versus confocal detection which we refer to as relative detection efficiency. The relative efficiency evidences the challenge to extract photons from a large detection volume with respect to a resolution limited spot. The waveguide transmission losses are disregarded in this comparison and will be considered later together with the absolute detection efficiency. The green curve marks the detection efficiency for a NV$^-$ sub-ensemble that is not perfectly centered compared to the waveguide mode (see fig. \ref{Fig:Fabrication}(c) green circle and dashed white cross for the excitation point), resulting in a reduction of the intensity by the factor 0.61 compared to the maximum at optimal position. We calculate relative detection efficiencies ranging from $\SI{8}{\percent}$ at $\SI{660}{\nano\metre}$ to more than $\SI{15}{\percent}$ at $\SI{735}{\nano\metre}$. The grey curve gives the efficiency for an ideally coupled NV$^-$-center located at the maximum of the waveguide mode, yielding ideal relative detection efficiencies up to $\SI{30}{\percent}$ at $\SI{735}{\nano\metre}$. The grey shaded area indicates all possible detection efficiencies from ideally to completely uncoupled ensembles. The black line marks the $1/e^2$-profile of the Gaussian mode that will be excited when operating the sensor through the waveguide. The figure illustrates the large sensing area of up to $\SI{105}{\micro\meter\squared}$ per waveguide mode with an average relative detection efficiency of $\SI{10.4}{\percent}$ for the large ensemble of NV$^-$-center which is addressed via the waveguide mode. 

We now quote the absolute detection efficiency taking into account the waveguide transmission losses including the outcoupling efficiency of up to $\SI{79.8}{\percent}$, or $\SI{6.95}{\decibel}$, which is comparable to values reported for type-II waveguide in diamond \cite{Sotillo2016, Sotillo2017}. The absolute detection efficiency of the confocal collection behind the objective is $\SI{2.5}{\percent}$ corresponding to an absolute detection efficiency of $\SI{0.05}{\percent}$ through the waveguides when averaged over the $1/e^2$-area of the Gaussian mode. The waveguide detection efficiency corresponds to a waveguide coupling efficiency of $\SI{0.3}{\percent}$ in agreement with estimated coupling efficiencies for individual NV$^-$-centers in laser-written waveguides \cite{Hadden2018} of $\SI{0.1}{\percent}$. In future designs, shorter waveguides and improved outcoupling performance will increase transmission, thus leading to higher detection efficiencies.

While the detection efficiency of a confocally excited NV$^-$-center ensemble is, as expected, decreased with respect to conventional confocal detection, the large number of addressable NV$^-$-centers via the waveguide mode increases the achievable overall sensitivity. We extrapolate the increased number of NV$^-$-centers that is accessible via the waveguide mode by comparing the mode field area of $\SI{105}{\micro\meter\squared}$ with the confocal spot size of $\SI{0.062}{\micro\meter\squared}$. Taking into account the homogeneous distribution of NV$^-$-centers over the entire waveguide mode, we expect an increase in the number of addressed NV$^-$-centers $N$ by a factor of 1690. Compared to conventional confocal detection, the sensitivity thus improves by a factor of 41 according to the $\sqrt{N}$-dependency. 

\section{Sensor performance}

\begin{figure*}[]
\includegraphics[scale=1.0]{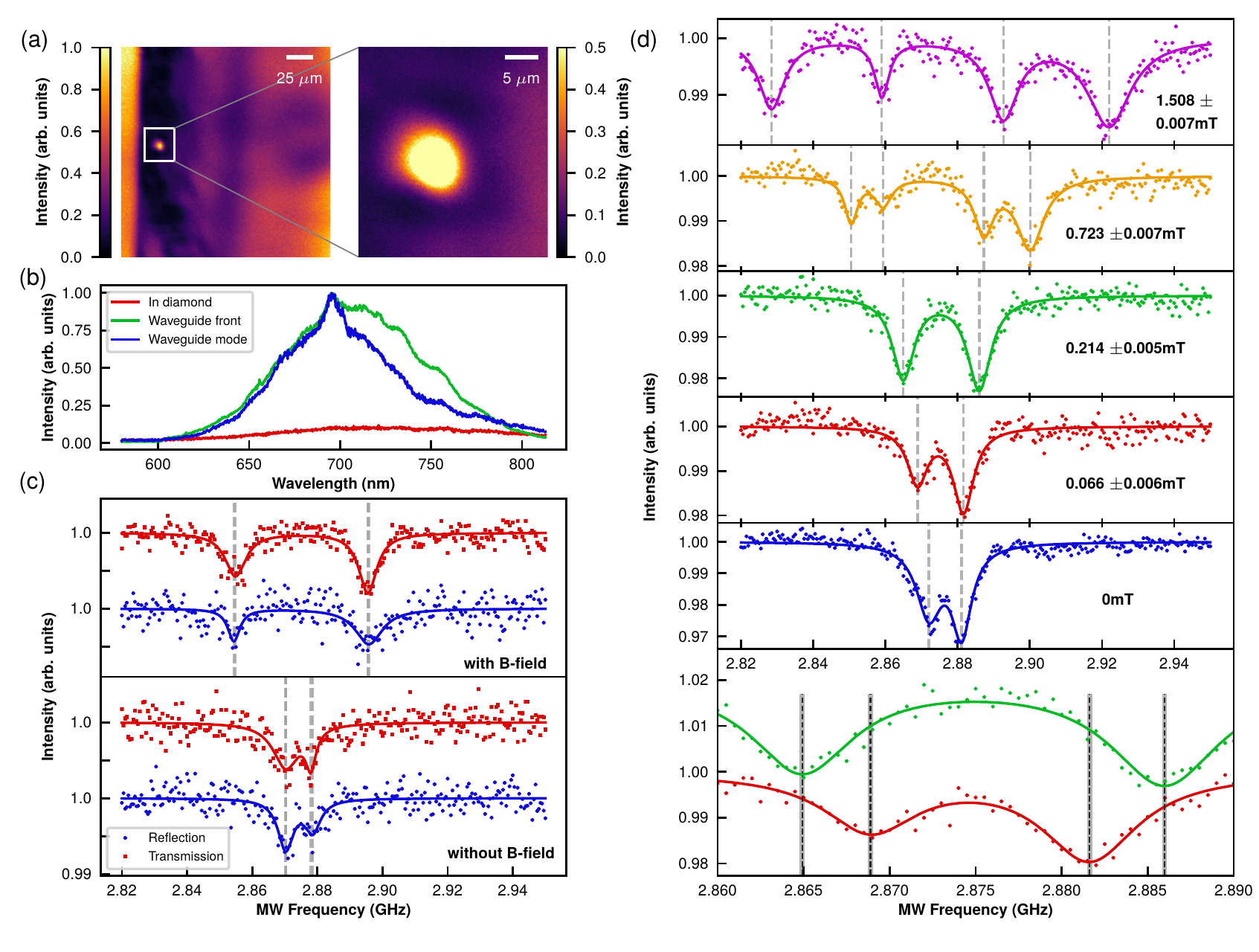}
\caption{\label{Fig:Backwards Transmission}(a) Confocal scan of the backside of the sample with a low-NA objective and a zoomed-in view of the waveguide mode. (b) Comparison of the PL spectra in transmission (blue) and reflection (green) of the waveguide mode as well as the transmitted signal outside the waveguide mode which we assign to the background signal (red). (c) Comparison of the ODMR spectrum at zero magnetic field and with applied magnetic field detected in confocal configuration (blue, reflection) and in transmission through the waveguide (red, transmission). (d) ODMR scans with varying magnetic field strengths between 0 mT and 1.508 mT. The lowest plot is a zoom-in comparing the ODMR scans at \SI{0.066}{\milli\tesla} and \SI{0.214}{\milli\tesla}. We infer the magnetic field sensitivity from the accuracy of the resonance frequency extracted from the Lorentzian fits.}
\end{figure*}

In the following, we study the sensing performance of our device. All measurements are performed at room temperature although the sensor can be operated over a large temperature range. We excite the NV$^-$-ensemble with off-resonant ($\SI{532}{\nano\metre}$) laser light, which is coupled from the back side into the waveguide. The NV$^-$-signal is also read out through the waveguide in order to keep the front facet of the diamond sensing area, which contains the shallow-implanted NV$^-$-ensemble, fully accessible. Figure \ref{Fig:Backwards Transmission}(a) shows a confocal scan of the diamond backside with the low-NA objective used for incoupling of excitation light and read-out of NV$^-$-signal through the waveguide. The waveguide mode is clearly resolved including the waveguide walls confining the mode, which appear dark. The spectrum of the waveguide-assisted emission resembles the spectrum of the NV$^-$-ensemble, as highlighted by comparison with a standard confocal spectrum of the ensemble in figure \ref{Fig:Backwards Transmission}(b). The background spectrum is shown in red  and is recorded spatially shifted from the waveguide mode. 

We now utilize the NV$^-$-ensemble as magnetic field sensor by exploiting its ODMR-signal. Therefore, we apply a MW field to the sample and vary an external magnetic field by approaching the sensor with a permanent magnet. Varying the MW frequency reveals the characteristic dips in the fluorescence signal, as shown in figure \ref{Fig:Backwards Transmission}(c). First, we compare the ODMR signal at zero magnetic field detected in standard confocal configuration with the transmission through the waveguide. Both signals reveal two characteristic dips separated by \SI{9}{\mega\hertz} due to strain induced by the laser-written waveguides \cite{Sotillo2018}, residual strain in the diamond lattice \cite{Kubo2010, Doherty2012} and also background magnetic field from the environment. An applied magnetic field increases the splitting between the two dips as shown in fig. \ref{Fig:Backwards Transmission}(c)), both in transmission and reflection. Figure \ref{Fig:Backwards Transmission}(d) shows the increased splitting with increasing magnetic field strength. At higher magnetic fields, each dip splits into two, since the NV$^-$-centers experience different magnetic fields depending on their orientation in the crystal lattice \cite{Matsuzaki2016}. We use the gyromagnetic ratio of an NV$^-$-center, $\gamma = \frac{g_\text{e} \mu_\text{B}}{\hbar} = \SI{28}{\giga\hertz\per\tesla}$, in order to calculate the splitting due to an applied magnetic field $B$ along the NV$^-$-center axis as $\Delta\Omega = 2 \gamma B_z$ \cite{Kuwahata2020}. From the spin Hamiltonian of the NV$^-$-center,
\begin{equation}
H = D\left(S_z^2 -\frac{1}{3} \left(S\left(S + 1\right)\right)\right) + E\left(S_x^2 - S_y^2\right) + g_\text{e} \mu_\text{B}\mathbf{B} \cdot \mathbf{S},
\end{equation}
we deduce the magnetic field \cite{Balasubramanian2008} as 
\begin{equation}
\left( g_\text{e} \mu_\text{B}B \right)^2 = \frac{1}{3} \left(\nu_1^2 + \nu_2^2 - \nu_1\nu_2 - D^2\right) - E^2.
\end{equation}
The zero field splitting parameter is given as $D$ and a nonzero $E$ incorporates additional splitting of the two dips, located at MW frequencies $\nu_1$ and $\nu_2$ at zero magnetic field due to strain. The zero-field splitting could be further increased by the earth magnetic field.
In our case, the ODMR measurements yield effective magnetic fields between \SI{0.066}{\milli\tesla} and \SI{1.508}{\milli\tesla} along the NV$^-$-center axis, respectively. 

Our measurements showcase the ability to measure magnetic fields with an accuracy better than \SI{6}{\micro\tesla} with our device. The lowest panel of figure \ref{Fig:Backwards Transmission}(d) illustrates the fitted peak positions together with their error margins that are now used to estimate the overall magnetic field sensitivity. The two measurements are performed at \SI{0.066}{\milli\tesla} and \SI{0.214}{\milli\tesla} applied magnetic field, respectively, with two clearly separated peak positions. Their fit error margins are still significantly separated. We calculate the error of the magnetic field difference of the two measurements including their fit errors. With these fit errors as reference, we estimate that magnetic field differences of \SI{6}{\micro\tesla} can be resolved with our device. The sensitivity $\eta_\text{DC}$ of our magnetometer device to DC magnetic fields for continuous ODMR measurements (CW-ODMR) can be calculated as \cite{Rondin2014}
\begin{equation}
\eta_\text{DC} = \frac{\Delta \nu}{\gamma C \sqrt{I_\text{PL}}},
\end{equation}
where $\Delta \nu$ denotes the full-width-half-maximum (FWHM) linewidth of the ODMR-dip, $C$ the ODMR-contrast and $I_\text{PL}$ the detected PL intensity in counts per second. In our typical \SI{30}{\minute} CW-ODMR measurements, we reach sensitivities up to \SI{36}{\micro\tesla\per\hertz\tothe{1/2}} in confocal configuration and \SI{62}{\micro\tesla\per\hertz\tothe{1/2}} when detecting the ODMR-signal through waveguide transmission, which correspond to FWHMs of \SI{7.5}{\mega\hertz} in both cases. Hereby, different count rates and PL contrasts lead to differing sensitivities. These sensitivities are similar to hybrid approaches creating integrated quantum sensors \cite{Ruan2018, Bai2020}. In the future, measuring the time-resolved photon noise on the point of maximal slope of the ODMR-signal will yield more accurate sensitivity estimation. Furthermore, the sensitivity can be increased when using pulsed ODMR or Ramsey spectroscopy techniques \cite{Levine2019}. Moreover, our measurement device works reliably over more than 20 hours and the minimal measurement times for a full MW frequency scan can be reduced to less than 40 seconds with our settings \footnote{Details see supplementary information}. 

\begin{figure}[]
\includegraphics[scale=1.0]{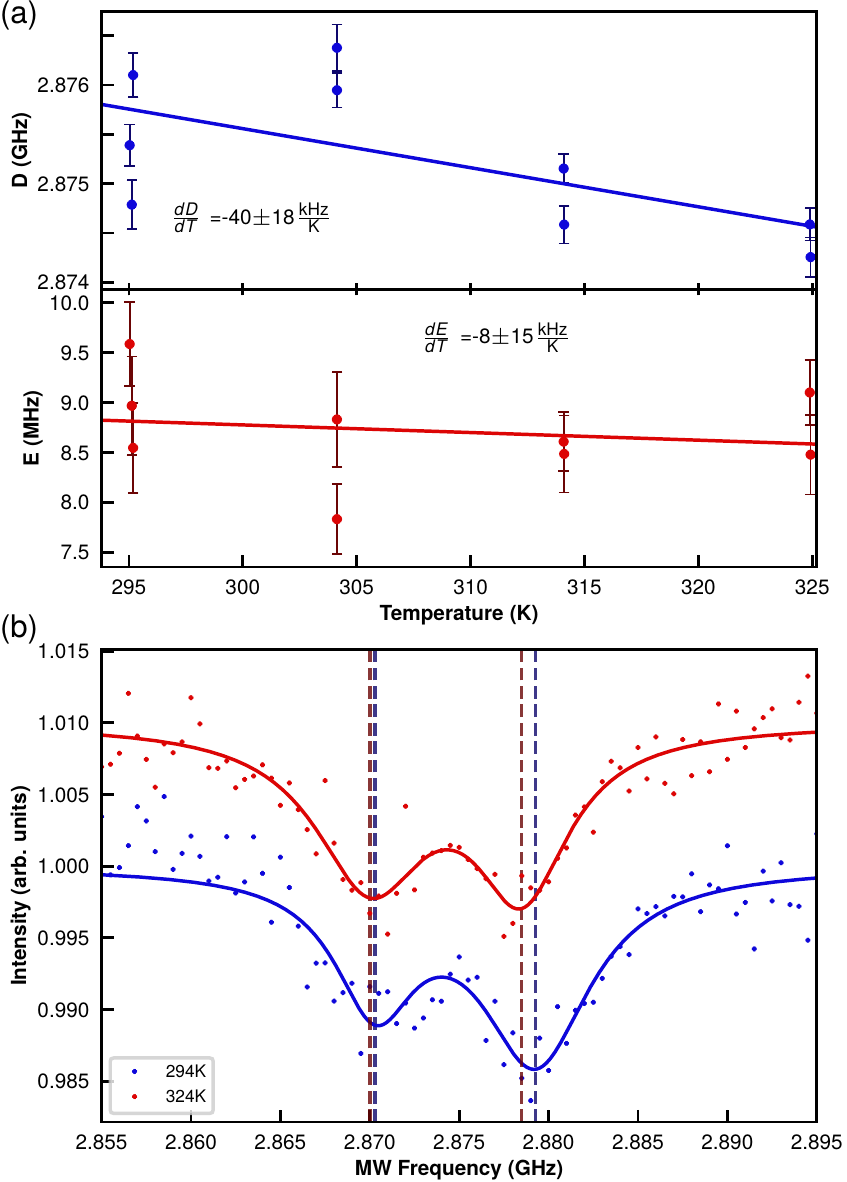}
\caption{\label{Fig:ODMR Tdependent}(a) T-dependence of the axial and transverse ZFS parameters $D$ and $E$. (b) ODMR scans at temperatures of \SI{294}{\kelvin} and \SI{324}{\kelvin} are used to illustrate temperature sensitivity of the ZFS parameters.}
\end{figure}

We now investigate the performance of our device as a temperature sensor by exploiting the temperature-dependence of the zero-field splitting (ZFS). Therefore, we heat the sample from room temperature (\SI{294}{\kelvin}) up to \SI{324}{\kelvin} and measure ODMR without applying an external magnetic field. To analyze the change of the axial and transverse ZFS parameters, $D$ and $E$, we follow studies of NV$^-$-ensembles in bulk diamond \cite{Acosta2010}. Linear fits to the data (see fig. \ref{Fig:ODMR Tdependent}(a)) reveal gradients of  $\frac{dD}{dT} = -40 \pm 18 \si{\kilo\hertz\per\kelvin}$ and $\frac{dE}{dT} = -8 \pm 15 \si{\kilo\hertz\per\kelvin}$, respectively. $\frac{dD}{dT}$ deviates from values in literature \cite{Acosta2010} by roughly a factor $1.5$. We tentatively ascribe this difference to a non-linear systematic temperature difference between the diamond sensor and the heater (together with the heat sensor) below the sample holder. Thereby, we systematically overestimate temperatures larger than \SI{294}{\kelvin}. However, the trend of $\frac{dE}{dT}$ agrees with literature studies \cite{Acosta2010}. A comparison of ODMR measurements at different temperatures underpins this finding (see fig. \ref{Fig:ODMR Tdependent}(b)). Note that the accessible temperature range for our device is only limited by the heating capabilities in our current setup and potentially covers the scale from high temperatures to cryogenic temperatures.

\section{Discussion and outlook}

Summarizing, we present a novel architecture for an integrated magnetometer based on NV$^-$-centers shallow-implanted in the field maximum of fs-laser-written type-II waveguides in diamond. We benchmark the waveguiding and coupling efficiency of the NV$^-$-centers in transmission as well as the performance of the platform as magnetic field and temperature sensor. We benchmark the magnetic field sensitivity to at least \SI{36}{\micro\tesla\per\hertz\tothe{1/2}}. Nevertheless, sensitivities below picotesla in bulk diamond \cite{Wolf2015, Fescenko2020} provide a perspective for future improvements regarding NV$^-$-center creation and ODMR detection techniques, for example operating at the highest sensitivity point (at the maximum slope of the ODMR line). The waveguide-assisted excitation and detection enables separation of the sensor operation from the sensing task making samples accessible which are fragile to light illumination, such as paramagnetic labels and biomolecules which are not photo-stable. The device is compatible with a large temperature regime from cryogenic temperatures to high temperatures. The access to cryogenic temperatures could prove versatile when studying open questions in material science such as superconductivity. 

Furthermore, due to the three-dimensional laser-writing capabilities \cite{Kononenko2011, Courvoisier2016, Gong2020}, the sensor can be extended to two-dimensional sensor arrays. The large area with accessible NV$^-$-centers can be used in the future to achieve large filling factors of two-dimensional sensing arrays. The $1/e^2$ mode area on the front facet covers about one half of the total waveguide area. Assuming densely packed lines of waveguides, where two horizontal waveguides share one laser written side-wall and two vertical waveguides are slightly spaced to avoid coupling, results in overall filling factors on the sensing surface of about 0.2. Each waveguide could be accessed individually and detect local temperatures and magnetic fields. The waveguide array could be operated simultaneously, thus allowing imaging of spatially resolved magnetic field distributions and their time dynamics in extension to probing magnetism \cite{Thiel2019} or antiferromagnetic order \cite{Gross2017} via scanning nanoprobes. Our sensing chips could take up a similar role such as CCD-camera chips with applications in spatially and temporally correlated measurement in material and life sciences. Furthermore, the combination with laser-written microfluidic channels \cite{Bharadwaj2019} could provide perspectives towards on-chip quantum sensing of liquids.

\section*{Methods}

\subsection*{Waveguide fabrication}

The femtosecond laser used for waveguide writing in diamond was a Yb:KGW Fiber Laser (Bluecut, Menlosystems) with 300-fs pulse duration, 515-nm wavelength (frequency doubled using an LBO crystal), focused with a 1.25-NA oil immersion lens (RMS100$\times$-O 100$\times$ Olympus Plan Achromat Oil Immersion Objective, 100$\times$ oil immersion, Olympus). Polished 2 mm $\times$ 2 mm $\times$ 0.3 mm synthetic single-crystal diamond samples (type II, electronic grade with nitrogen impurities $<5$ ppb) were used. The sample was placed on a computer-controlled, 3-axis motion
stage (ANT130 series, Aerotech) to translate the sample relative to the laser to form the desired photonic structures.  The polarization of the incident laser was perpendicular to the scan direction.

\subsection*{NV$^-$-center creation}

The NV$^-$ sensor used in this work was fabricated by ion implantation on the front facet of a commercial type IIa diamond substrate. Using a home built low energy ion implanter equipped with a Wien mass filter as well as an Einzel lens for beam focusing, nitrogen ions of energy 5 keV were implanted at a dose of $5 \times 10^{11}~\mathrm{\frac{^{15}N^+}{cm^2}}$ and a beam diameter of approximately \SI{100}{\micro\meter}. Subsequently, the substrate was annealed in ultra-high vacuum (UHV) for \SI{3}{\hour} at $1000~^{\circ}\mathrm{C}$ to form NV$^-$-centers. The ramp up time was $7~\mathrm{\frac{^{\circ}C}{min}}$ with an intermediate soak for \SI{1}{\hour} at $500~^{\circ}\mathrm{C}$. Between all processing steps, the substrate was boiled several hours in a 1:1:1 mixture of sulphuric, perchloric and nitric acid to remove any organic or graphitic residues from the surfaces of the diamond. Details on the annealing and ion implantation setups for sensor fabrication can be found in \cite{Lang2020}.

\subsection*{Confocal setup}

The sample was mounted in a liquid helium, continuous flow cryostat enabling operation temperatures down to $\SI{5}{\kelvin}$. The temperature of the sample could further be varied and controlled with a heater which was in thermal contact with the device. The waveguides could be accessed via a high-NA (0.9) objective from one side of the diamond slab containing the NV$^-$-centers and from the backside with a low-NA (0.25) objective to collect the transmission through the waveguide. The objectives were mounted on movable stages for optimization of the in- and outcoupling. Thereby, we combined a standard confocal setup in front with additional waveguide transmission detection at the back of the sample. The characterization of the emitters could therefore be done in standard confocal spectroscopy. For measuring waveguide losses in our setup, we coupled light through the waveguides with a second low-NA objective and detect light transmitted through the waveguide at the back. The NV$^-$-center ensemble could also be excited directly with a high-NA objective while simultaneously reading out through the waveguide with low-NA objective. For CW-ODMR measurements we spanned a wire with $\SI{20}{\micro\meter}$ diameter close to the NV$^-$ ensemble to deliver the MW signal, which were generated by a vector signal generator (Rohde \& Schwarz SMIQ 03). Typical CW-ODMR scans in our setting took \SI{30}{\minute} to ensure good signal to noise. A single measurement consisted of many repeated quick scans (around \SI{7}{\second}) over a chosen frequency range. We then averaged over the ensemble of fast scans to get a result for the \SI{30}{\minute} CW-ODMR scan.

\section*{Acknowledgments}

\medskip
\begin{acknowledgments}
The authors would like to thank Felix Breuning for experimental support in the beginning of the project. IFN-CNR is grateful for technical assistance from Erasmus students Marie Dartiguelongue and Héloïse Raugel. IFN-CNR and UUlm are grateful for support from the H2020 Marie Curie ITN project LasIonDef (GA n.956387). A.K. acknowledges support of the European Regional Development Fund (EFRE) program Baden-Württemberg. M.K.K. and A.K. acknowledge support of IQst. M.H. acknowledges support from the Studienstiftung des deutschen Volkes. V.B. is thankful for financial support from the ERC project PAIDEIA GA n.816313. A. G. is thankful for support from the Lombardy Region Project sPATIALS3, cofunded by POR FESR 2014-2020 Call HUB Ricerca e Innovazione and the H2020 Marie Curie ITN project PHOTOTRAIN (GA n.722591).
\end{acknowledgments}

\section*{Author Contributions}
M.H., M.K.K., V.B., S.M.E and A.K. conceived the project. M.H. and M.K.K. performed and analyzed all PL and ODMR measurements. V.B., J.P.H., R.Y., A.N.G., R.R. and S.M.E. produced the laser-written WG in diamond, which were implanted and annealed by J.L. and F.J. to create NV$^-$-centers. All authors discussed the results. M.H., M.K.K. and A.K. wrote the manuscript, which was discussed and edited by all authors.

\bibliography{WG-NVTransmission}

\end{document}